\documentstyle[sprocl]{article}

\input{psfig}

\def\NPB#1#2#3{{\em Nucl.~Phys.} B {\bf #1}, #3 (#2)}
\def\PLB#1#2#3{{\em Phys.~Lett.}  B {\bf #1}, #3 (#2)}
\def\PRL#1#2#3{{\em Phys.~Rev.~Lett.} {\bf #1}, #3 (#2)}
\def\PRD#1#2#3{{\em Phys.~Rev.} D {\bf #1}, #3 (#2)}
\def\prep#1#2#3{{\em Phys.~Rep.} {\bf #1}, #3 (#2)}

\def\ra{\rightarrow}

\def\be{\begin{equation}}
\def\ee{\end{equation}}
\def\bea{\begin{eqnarray}}
\def\eea{\end{eqnarray}}

\newcommand{\chim} [1] {\tilde{\chi}^{-}_{#1} }

\newcommand{\chipm} [1] {\tilde{\chi}^{\pm}_{#1} }
\newcommand{\chiz} [1] {\tilde{\chi}^{0}_{#1} }
\newcommand {\glu} {\tilde{g} }

\newcommand {\mglu} {m_{\tilde{g}} }

\newcommand{\tanb} {\tan \beta}

\def\lsim{\raise0.3ex\hbox{$\;<$\kern-0.75em\raise-1.1ex\hbox{$\sim\;$}}}
\def\gsim{\raise0.3ex\hbox{$\;>$\kern-0.75em\raise-1.1ex\hbox{$\sim\;$}}}

\begin{document}

\title{GLUINO CASCADE DECAYS WITH BROKEN R--PARITY}

\vspace*{-2.5cm}
\begin{flushright}
  UWThPh-1997-50 \\
  HEPHY-PUB 683/97 \\
  hep-ph/9712484
\end{flushright}
\vspace*{0.6cm}

\author{A. BARTL$^*$, F. DE CAMPOS$^\dag$, M. A. GARC\'IA-JARE\~NO$^\ddag$,
       M. B. MAGRO$^\P$, \\ W. MAJEROTTO$^\diamond$, 
       \underline{W. POROD}$^{*,a}$, J. W. F. VALLE$^\ddag$}

\address{$*$ Institut f\"ur Theoretische Physik, University of Vienna, 
            A--1090 Vienna, Austria \\ 
         $\dag$ Instituto de F\'{\i}sica Te\'orica -- Universidade Estadual
            Paulista  Rua Pamplona, \\ 145 -- 01405-900 -- S\~ao Paulo -- SP,
            Brasil \hfill \\ 
         $\ddag$ Instituto de F\'{\i}sica Corpuscular -- C.S.I.C.
            Departament de F\'{\i}sica Te\`orica, Universitat de Val\`encia
            46100 Burjassot, Val\`encia, Spain \hfill \\
         $\P$ Dto. Fisica Matematica, Uni. Sao Paulo.
            Rua do Matao S/N -- Cid. Universitaria
            C.P. 66318 CEP 05315-970 Sao Paulo, Brasil \hfill \\
         $\diamond$ Institut f\"ur Hochenergiephysik, Akademie der 
            Wissenschaften, A--1050 Vienna, Austria}


\maketitle\abstracts{ 
We study the pattern of gluino cascade decays in a class of supersymmetric 
models where R--parity is spontaneously broken. The multi--lepton 
and same--sign dilepton rates in these models are compared with 
those of the Minimal Supersymmetric Standard Model.
We show that these rates can be substantially enhanced in 
models with broken R--parity.}

\footnotetext[1]{ Talk presented at the {\it International Workshop on 
Physics Beyond The Standard Model: from Theory to Experiment}, 
October 13 -- 17, 1997,  Val\'encia, Spain}

\section{Introduction}

The search for supersymmetric particles will be one of the main topics in the
experimental program of LHC. At LHC gluinos and squarks with masses
up to $\lsim 1.5$~TeV can be discovered \cite{aachen}. 
Most studies of gluino production and decays \cite{aachen,barnett}
have been performed in
the Minimal Supersymmetric Standard Model (MSSM)\cite{mssm}. 
A characteristic feature of MSSM is the
conservation of R--parity implying the stability of the lightest
supersymmetric particle (LSP). 
Effects of R--parity breaking have been explored in \cite{binet,rpglu}.
In this contribution we study the impact of spontaneous R--parity  breaking on
gluino cascade decays \cite{rpglu}.

R--parity can be broken spontaneously
through non--zero vacuum expectation values 
for scalar neutrinos \cite{romao}. 
There are two generic cases of models with spontaneous R--parity breaking.
Firstly, if lepton number is part of the gauge symmetry there is a new
gauge boson $Z^\prime$ which gets mass via the Higgs mechanism
\cite{ZR}. In this model
the LSP is in general a neutralino which
decays mostly into Standard Model fermions.
Secondly, if spontaneous R--parity violation occurs
in the absence of any additional gauge symmetry, it leads to the
existence of a physical massless Nambu--Goldstone boson, called
majoron (J). 
As a consequence the lightest
neutralino $\chiz{1}$ may decay invisibly as
$\chiz{1} \ra \nu + J$.
 
We also consider a specific class of models
with explicit R--parity breaking characterized by a single 
bilinear superpotential term of the type $\ell H_u$
\cite{epsi}. These models mimic in many 
respects the features of models with spontaneous breaking 
of R--parity containing an additional gauge boson. 
In the following the class of models containing a majoron will 
be denoted by J--model, whereas the models without a 
majoron will be denoted by $\epsilon$--model \cite{epsi}.

\section{Gluino Cascade Decays}

In the following we
will assume that squarks are heavier than gluinos, so that pair
production of gluinos dominates. As an example we consider a  
gluino with a mass of 500 GeV. At LHC with a centre of mass energy 
of 14 TeV the production cross section will be $\sim 25$ pb, which 
corresponds to 2.5~$10^6$ gluino--pairs per year for an integrated 
luminosity of $10^5$~pb$^{-1}$.
The gluino has the following decays \cite{Bartl91}:
$\glu \to q\bar{q}\chiz{i}$, $q\bar{q}'\chipm{j}$, $g \chiz{i}$,
where $\chiz{i}$ denotes the neutralinos and $\chipm{j}$ the charginos.
The charginos and neutralinos decay further, giving rise to cascade decays.
In models with spontaneous R--parity breaking one has additional decay
modes for neutralinos and charginos compared to the MSSM. They are induced by 
the mixing between charged leptons and charginos, and between
neutrinos and neutralinos \cite{rpglu,MASIpot3}.

Among the various signals of gluinos 
the multi--lepton (ML) and the same--sign dilepton (SSD)
signals are experimentally very important. 
We calculate the rates for the ML and SSD signals in gluino pair 
production for MSSM, the J--model and the $\epsilon$--model. 
We count all leptons coming from charginos, neutralinos, 
$t$--quarks, $W$-- and $Z$--bosons, summing over electrons and muons. We 
take $\mglu = 500$~GeV, $\tanb = 2$,
$M_2 = 170$~GeV, $m_A = 500$~GeV.
For the parameters characterizing R--parity violation we take
$h_{\nu 33} = 0.01$, $v_{R3} = 100$~GeV and
$v_{L 3} = 10^{-5}$~GeV. In the $\epsilon$--model
this corresponds to $\epsilon = 1$~GeV. 
The $\mu$ parameter is varied between 
$-$1~TeV and 1~TeV.
We take into account the restrictions on these parameters that follow 
from searches for SUSY particles at LEP \cite{magro} and 
at TEVATRON \cite{D0}. Moreover, we fulfil
the constraints from  neutrino physics and 
weak interactions phenomenology \cite{bounds} to
which R--parity breaking models are sensitive.

\begin{figure}[t!]
{\setlength{\unitlength}{1mm}
\begin{picture}(115,124)
\put(-2,-20){\mbox{\psfig{figure=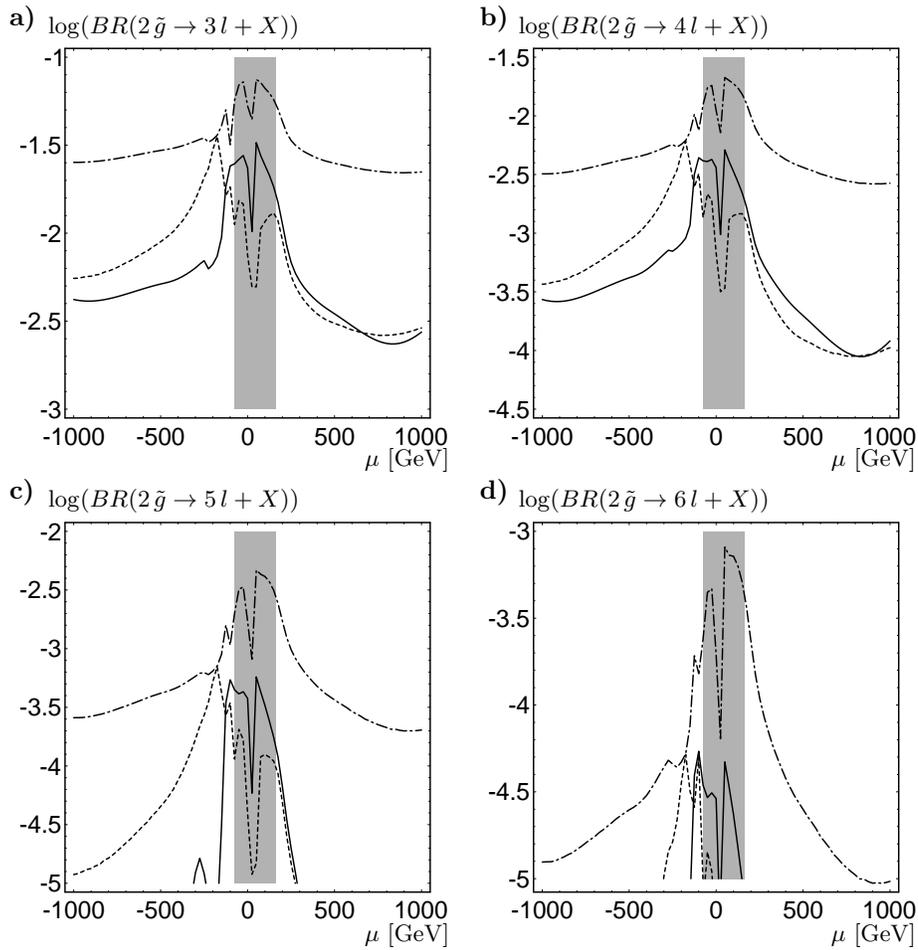,height=16cm}}}
\put(-0.5,121){\makebox(0,0)[bl]{{\bf a)}}} 
\put(3.5,120){\makebox(0,0)[bl]{
    {\small $\log ( BR(2 \, \glu \to 3 \, l + X))$}}}
\put(58, 62.5){\makebox(0,0)[br]{{\small $\mu$~[GeV]}}}
\put(-0.5,58){\makebox(0,0)[bl]{{\bf c)}}} 
\put(3.5,57){\makebox(0,0)[bl]{
    {\small $\log ( BR(2 \, \glu \to 5 \, l + X))$}}}
\put(58,-0.8){\makebox(0,0)[br]{{\small $\mu$~[GeV]}}}
\put(62,121){\makebox(0,0)[bl]{{\bf b)}}} 
\put(66,120){\makebox(0,0)[bl]{
    {\small $\log ( BR(2 \, \glu \to 4 \, l + X))$}}}
\put(120.0,62.5){\makebox(0,0)[br]{{\small $\mu$~[GeV]}}}
\put(62,58){\makebox(0,0)[bl]{{\bf d)}}} 
\put(66,57){\makebox(0,0)[bl]{
    {\small $\log ( BR(2 \, \glu \to 6 \, l + X))$}}}
\put(120.0,-0.8){\makebox(0,0)[br]{{\small $\mu$~[GeV]}}}
\end{picture}}
\caption{
Multi--lepton signals (summed over electrons and muons) as a 
function of $\mu$. The parameters are given in the text.
We show a) the 3--lepton, b) the 4--lepton, 
c) the 5--lepton and d) the 6--lepton signal for the MSSM (full line), 
the J--model (dashed line) and the $\epsilon$--model (dash--dotted 
line). The  shaded area is covered by LEP2.
}
\end{figure}

\begin{figure}[t!]
{\setlength{\unitlength}{1mm}
\begin{picture}(119,66)
\put(1,0){\mbox{\psfig{figure=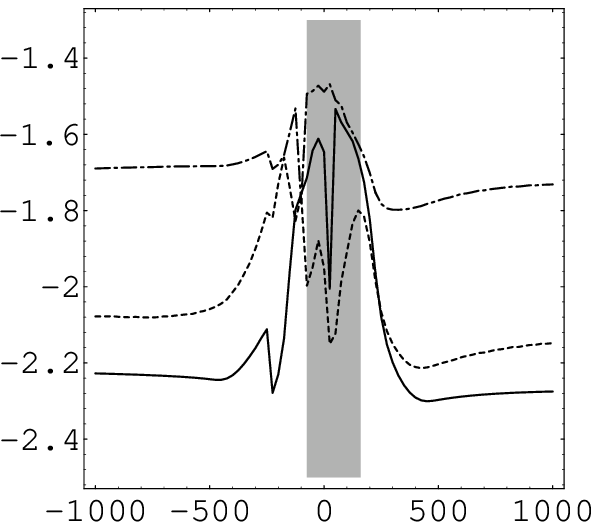,height=6.5cm}}}
\put(0.5,63){\makebox(0,0)[bl]{
    {\small $\log ( BR(2 \, \glu \to l^+ l^+, \, l^- l^- ))$}}}
\put(67, 1.2){\makebox(0,0)[br]{{\small $\mu$~[GeV]}}}
\put(70,63){ \begin{minipage}[t]{4.7cm}
\vspace*{11mm} 
\caption{
The same--sign dilepton signal (summed over electrons and muons)
as a function of $\mu$. The parameters are given in the text.
Predictions in MSSM (full line), 
J--model (dashed line) and $\epsilon$--model (dash--dotted line).
The shaded area is covered by LEP2.}
             \end{minipage} }
\end{picture}}
\end{figure}

In Fig.~1 we show the branching ratios for the 3--, 4--, 5-- and 
6--lepton events.
Quite generally, the various ML rates in the R--parity violating models
can be different from those in the MSSM for two reasons:
(i) The lightest neutralino $\chiz{1}$ can decay leptonically 
as $\chiz{1} \to Z^{(*)} \nu_{\tau} \to l^+ l^- \nu_{\tau}$, 
$\chiz{1} \to W^{(*)} \tau \to l^+ \nu_l \tau$, leading to an 
enhancement of the multi--lepton rates. 
(ii) The R--parity violating decays of the lightest chargino
$\chipm{1}$ and the second lightest neutralino $\chiz{2}$ may 
reduce the leptonic signal, $\chiz{2} \to W^{(*)} \tau$, 
$J \nu_{\tau}$, $\chim{1} \to J \tau$
(we do not count $\tau$ as a lepton).
Depending on which of these two effects is dominant, one has an
overall enhancement or a reduction of the leptonic rates compared 
to those expected in the MSSM. 
Note that even the 
6-lepton signal has a rate up to $5 \times 10^{-5}$ in the range
$-300$~GeV$< \mu < -80$~GeV, giving 125 events per year.

In Fig.~2 we show the SSD signal which
is enhanced in the J--model for  
$\mu \lsim -100$ GeV or $\mu \gsim 200$ GeV. This is due to the fact that
at least one of the neutralinos has a sizeable branching ratio into a 
$W$, leading to the enhancement of the signal.
In the $\epsilon$--model the signal is larger by an order of magnitude 
except for $|\mu| \lsim 200$~GeV. 

\section{Conclusions}

The effects of R--parity violation in gluino cascade decays have been studied
for two different classes of models, the J--model and the
$\epsilon$--model. We have calculated the rates for the ML and SSD signals. 
These processes are interesting from the experimental point of view since 
the 4--, 5--, 6--lepton signals are practically free of 
background from Standard Model processes. 
Comparing the J--model with MSSM, the ML and SSD signals can
increase or decrease depending on the model parameters, whereas
in the $\epsilon$--model all signals are enhanced
by one order of magnitude for most of the parameter ranges considered.

\section*{Acknowledgements}

W. P. thanks Prof. J. W. F. Valle and the organizers for the invitation
to this interesting and inspiring workshop.
This work was supported by 
{\sl Fonds zur F\"orderung der wissenschaftlichen Forschung}, 
Projekt Nr. P10843--PHY and 
by Acci\'on Integrada Hispano--Austriaca (''B\"uro f\"ur 
technisch-wissen\-schaft\-liche Zusammenarbeit des \"OAD'',
project no. 3).

\end{document}